\documentclass[aps,prl,twocolumn,noshowpacs,superscriptaddress,floatfix]{revtex4-1}
\usepackage{amsmath}
\usepackage{amssymb}
\usepackage{CJK}
\usepackage{graphicx}
\usepackage{dcolumn}
\usepackage{bm}
\usepackage{color}
\usepackage{ulem}
\usepackage{booktabs}
\usepackage{epstopdf}
\usepackage{natbib}
\begin{document}
\begin{CJK*}{GB}{gbsn}
\title{Reaction rate weighted multilayer nuclear reaction network}
\footnotetext{\hspace*{-5.4mm}Supported by the National Natural Science Foundation of China (Contracts Nos.~11890714, 1421505, 11875133 and 11075057), the National Key R{\&}D Program of China (Grant No. 2018YFB2101302), the Key Research Program of Frontier Sciences of the CAS (Grant No.~QYZDJ-SSW-SLH002), and the Strategic Priority Research Program of the CAS ( Grant No.~XDB34030200).

\noindent$^{*}$Corresponding authors. Email: ddhan@fudan.edu.cn; mayugang@fudan.edu.cn
}

\author{H. L. Liu}
\affiliation{Key Laboratory of Nuclear Physics and Ion-beam Application (MOE), Institute of Modern Physics, Fudan University, Shanghai 200433, China}
\affiliation{Shanghai Institute of Applied Physics, Chinese Academy of Sciences, Shanghai 201800, China}

\author{D. D. Han}\thanks{ddhan@fudan.edu.cn}
\affiliation{School of Information Science and Technology, Fudan University, Shanghai 200433, China}

\author{Peng Ji}
\affiliation{The Institute of Science and Technology for Brain-inspired Intelligence (ISTBI), Fudan University, Shanghai 200433, China}

\author{Y. G. Ma}\thanks{mayugang@fudan.edu.cn}
\affiliation{Key Laboratory of Nuclear Physics and Ion-beam Application (MOE), Institute of Modern Physics, Fudan University, Shanghai 200433, China}

\date{\today}

\begin{abstract}
Nuclear reaction rate ($\lambda$)  is a significant factor in the process of nucleosynthesis. A multi-layer directed-weighted nuclear reaction network in which  the reaction rate as the weight, and neutron, proton, $^4$He and the remainder nuclei as the criterion for different reaction-layers is for the first time built 
based on all thermonuclear reactions in the JINA REACLIB database.  Our results show that with the increase of the stellar temperature ($T_{9}$), the distribution of nuclear reaction rates on the $R$-layer network demonstrates a transition from unimodal  to bimodal distributions.
 Nuclei on the $R$-layer in the region of $\lambda = [1,2.5\times10^{1}]$ have a more complicated out-going degree distribution than the one in the region of $\lambda = [10^{11},10^{13}]$, and 
the number of involved nuclei at $T_{9} = 1$ is very different  from the one at $T_{9} = 3$. The redundant nuclei in the  region of $\lambda = [1, 2.5\times10^{1}]$ at $T_{9} = 3$ prefer  $(\gamma,p)$ and $({\gamma,\alpha})$ reactions to the ones at $T_{9}=1$, which produce nuclei around the $\beta$ stable line. This work offers a novel way to the big-data analysis on nuclear reaction network at stellar temperatures.

\end{abstract}


\maketitle



The mechanism of the nucleosynthesis and nuclear astrophysics process has attracted a great interest.
\textsuperscript{\cite{burbidge1957,schatz1998,arnould2007,kappeler2011,Chen,ji2016,r1,Fynbo,An2,NST_Pais,NST_Tang,NST_Ma,NST_Li,CPL1,CPL2}} Nuclear reaction rate, as an important input quantity in the calculation of the nuclear astrophysics network, can determine the path of nuclear reactions, and it can further affect the process of stellar evolution in the Universe and nuclear landscape.\textsuperscript{\cite{KH2015,PK2001,landscape1,landscape2}} Precise measurements of nuclear reaction rates  as well as neutron-, proton- and photo-induced reaction cross sections 
 have been intensively investigated in nuclear astrophysics 
\textsuperscript{\cite{DNS2003,SDSS,PDS2004,nA1,nA2,nA3,pA1,pA2,pA3,Photo1}} as well as superheavy element synthesis.\textsuperscript{\cite{SHE1,SHE2,SHE3,SHE4,SHE5,SHE6}} Additionally, features of $\alpha$-clustering\textsuperscript{\cite{cluster0,cluster1,cluster2,cluster3,cluster4}} and other  exotic structures of light nuclei \textsuperscript{\cite{exo1,exo2,exo3,exo4,exo5,exo6}} as well as  the determination of  mass and binding energy\textsuperscript{\cite{Eb0,Eb1,Eb2,Eb3,Eb4,Eb5}} are related to nuclear astrophysics process. 
Several massive nuclear datasets, including  REACLIB\textsuperscript{\cite{RHC2010}} and NACRE\textsuperscript{\cite{CM1999-2,Xu-NACRE-II}} databases,  have been constructed to facilitate this research topic. 
REACLIB database contains  information about different reactions and the corresponding reaction rate parameters, maintained by the Joint Institute for Nuclear Astrophysics (JINA). In this work,  we use the data from REACLIB V2.0, which includes 8048 nuclei and 82851 nuclear reactions, in which the detailed information related to reaction rate consisting of reaction types, reactants and products, and $Q$-value are provided.\textsuperscript{\cite{RHC2010}}

On the other hand, complex network science has achieved significant advances in recent years.\textsuperscript{\cite{BA1999-19,BS2006-20,CL2011-21,road}} Various real systems, such as internet, social connections, epidemics spreading, and chemical reactions etc. can be treated as complex networks to facilitate  investigations.\textsuperscript{~\cite{WS1998,JCC2010,JCC2012,Han1,Han2,Qian1,Qian2}} The main idea of the  complex network construction is to  consider  units as nodes and the interaction between two units as an edge. Furthermore,  many features of real systems can be investigated by taking advantage of the topological characteristics of the network, and this can further uncover some  hidden properties. 
In our previous work, using the REACLIB, we  considered four layers ($N$-layer, $P$-layer, $H$-layer and $R$-layer), and then formulated a multi-layer directed non-weighted nuclear reaction networks via the substrate-product method, 
moreover, solely studied its topological features.\textsuperscript{~\cite{ZL2016}}
The reaction rate, however, is an important input quantity in the process of nucleosynthesis,  which remains to be taken into account during the construction of the nuclear reaction network.

Previous nuclear astrophysics studies mainly concentrated on the precise calculation of nuclear reaction rates involve different types of nuclear reactions.\textsuperscript{\cite{CM1999-2,Guo2006}} In this Letter, however, we  focus on the topological characteristics of the directed-weighted nuclear reaction network with the reaction rate as the weight and particularly conduct statistical analysis of the reaction path of the nuclei.

\label{illustration}

The reactions in the REACLIB database, in general, are reversible and contain forward and backward directions. For the forward reaction, the reaction rate  $\lambda$ can be calculated by a parametrized function, which contains seven parameters $a_{0}$-$a_{6}$ shown as follows {\textsuperscript{~\cite{RT2000}}}:
\begin{equation}\label{rateequ}
    \lambda = \text{exp}(a_{0} + \sum_{i=1}^{5}{a_{i}T_{9}^{\frac{2i-5}{3}}} + a_{6}{\text{ln}}T_{9})
\end{equation}
where $a_{0}\sim a_{6}$ are given in the REACLIB database and the $T_{9}$ given in a unit of $10^9$K represents stellar temperature  where nuclei take part in different reactions.Although, these parameters for the backward reaction are not presented directly in the REACLIB database, the reverse reaction rate can be computed with the help of the partition functions in the file of WINVN.\textsuperscript{~\cite{RT2000,RT2001-1}}  The value of reverse reaction rate is equal to the forward reaction rate times the partition functions. In general, the fitted reaction rates are only valid in region of some temperatures, and REACLIB database provides 24 temperatures $T_9$ including 0.1, 0.15, 0.2, 0.3, 0.4, 0.5, 0.6, 0.7, 0.8, 0.9, 1.0, 1.5, 2.0, 2.5, 3.0, 3.5, 4.0, 4.5, 5.0, 6.0, 7.0, 8.0, 9.0, and 10.0. In the present manuscript, we only discuss the results at $T_9$ = 0.1, 1 and 3. 

In our previous directed non-weighted nuclear reaction network based on REACLIB and the substrate-product method, 
all nuclear reactions were divided to four types of $N$-layer, $P$-layer, $H$-layer and $R$-layer, which denote, respectively, the reactant of the neutron, proton, $^4$He and reminders. 
For each nucleus (or `node' $X$ in terms of network concept), we can identify  whether the $N$-layer reaction is the `in-coming'  (e.g $n$ + $Y \rightarrow X$), or `out-going' ($X + n \rightarrow  Y$).  The number of reactions for each nucleus (node) is represented by the number of `degree' which is  divided into either  in-coming degree or out-going degree. 
With this definition, the in-coming and out-going degree  distributions of nuclei  in the nuclear landscape can be accumulated for  each layer of nuclear reaction network. In this Letter, we establish a directed-weighted nuclear reaction network where nuclear reaction rate is taken as the weight of edges, which is based on the following consideration: 
the existence of edges of the nuclear reaction network  indicates that a certain  reaction  can take place, and the magnitude of nuclear reaction rate  quantifies the reaction probability of this reaction, and the corresponding edge is more important  in the weighted network.

Considering that the $R$-layer network has more complicated structure in comparison with  the $N$-layer, $P$-layer and  $H$-layer
in a directed non-weighted nuclear reaction network ~\cite{ZL2016,Liu1,Liu2,Liu3}, 
here we also focus the $R$-layer topology in the framework of directed-weighted network. In what follows, we perform the analysis on $R$-layer network  in which the nuclei mainly involve in photodisintegration reactions and $\beta$ decay reactions process under different temperature $T_{9}$.

\begin{figure}
\vspace{-1.5cm}
\includegraphics[width=8.6cm]{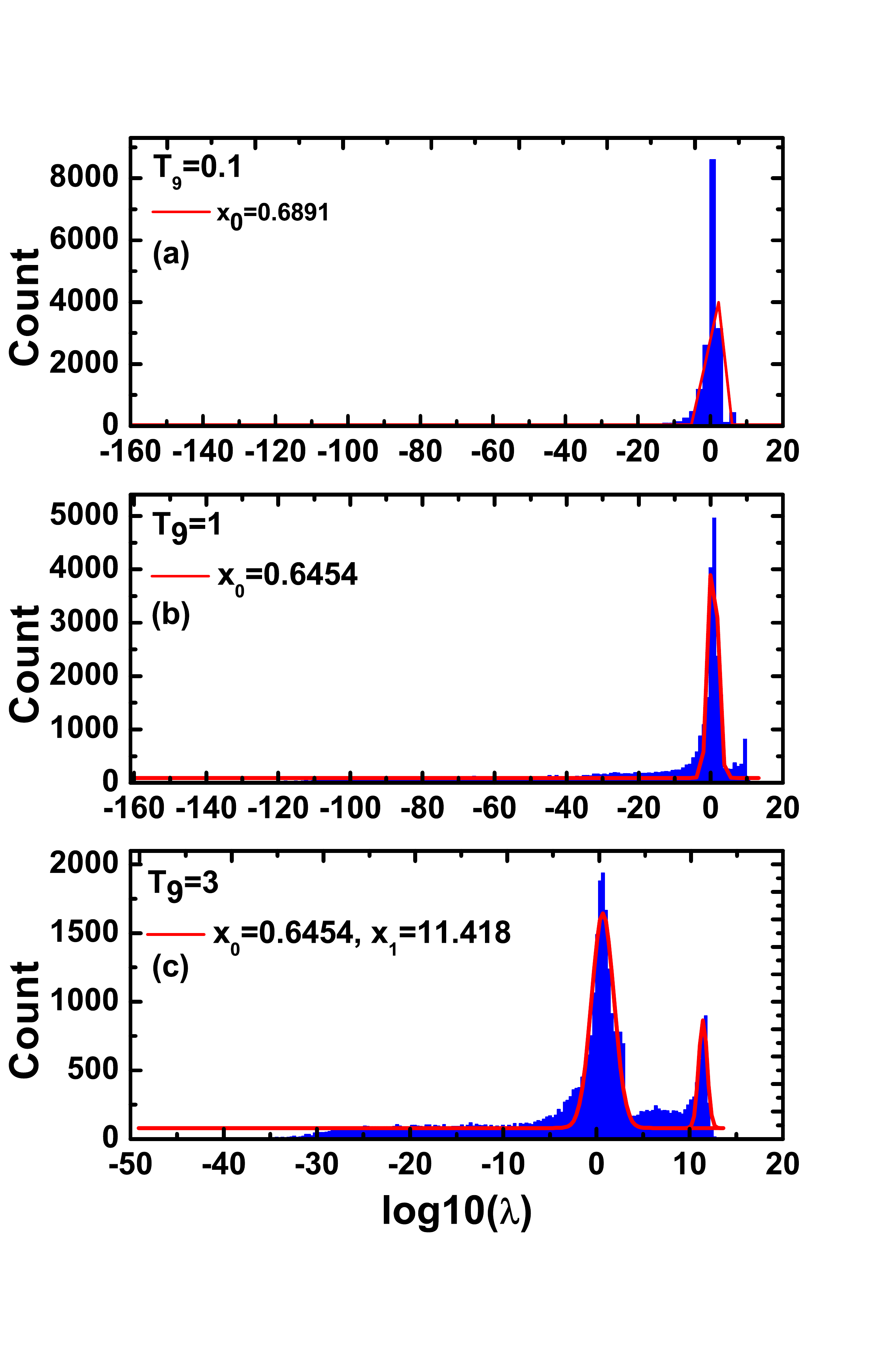}
\vspace{-1.5cm}
\caption{Distribution of reaction rates of all nuclei with the Gaussian fits in the $R$-layer at three different $T_{9}$: 0.1 (a),  1.0  (b),  and 3.0 (c).  From the fits, it indicates the positions of the peaks for $T_{9}$ = 0.1 and 1.0 are $\lambda = 10^{0.6891}$ and $10^{0.6454}$, respectively, while  positions of the double peaks for  $T_{9} = 3$ are $\lambda = 10^{0.6714}$ and $10^{11.418}$.}
\label{fit}
\end{figure}

On the basis of the reaction rate equation~\eqref{rateequ}, each  nuclear reaction rate in the $R$-layer network can be obtained at different temperatures combining with the REACLIB and WINVN databases. We found that the reaction rate has a heterogeneous distribution, ranging from $10^{-100}$ to $10^{50}$. For an example, Fig.~\ref{fit} demonstrates such distribution of nuclei in the $R$-layer network at three values of $T_{9}$ = 0.1, 1.0 and 3.0. From the perspective of stellar element burning process, the above temperatures are   close to the threshold  of He- (0.15 $\sim$ 0.23 $T_9$) \textsuperscript{\cite{He}}, Ne- (1.4 $\sim$ 1.7 $T_9$) \textsuperscript{\cite{Ne}} and Si- ( $\sim$ 3.0 $T_9$) \textsuperscript{\cite{Si}} burning processes. 
Figure~\ref{fit} indicates that the distribution of reaction rates of all nuclei varying from a unimodal distribution to a bimodal distribution with the increases of $T_{9}$  from 0.1 to 3.0. This results in many questions: does the  second peak at high temperatures come from the same nuclei as for the first peak but due to the increasing of nuclear reaction rate, or come from other nuclei not for the first peak?   Whether are the nuclei in the regions of the two peaks same or the same type of reactions? What is the degree distribution of the nuclei in the region of two peaks at different temperatures? In order to address these questions, we perform Gaussian fits to the peak distributions 
shown in  Fig.~\ref{fit}.

\begin{figure}[!htbp]
   \includegraphics[width=8.0cm]{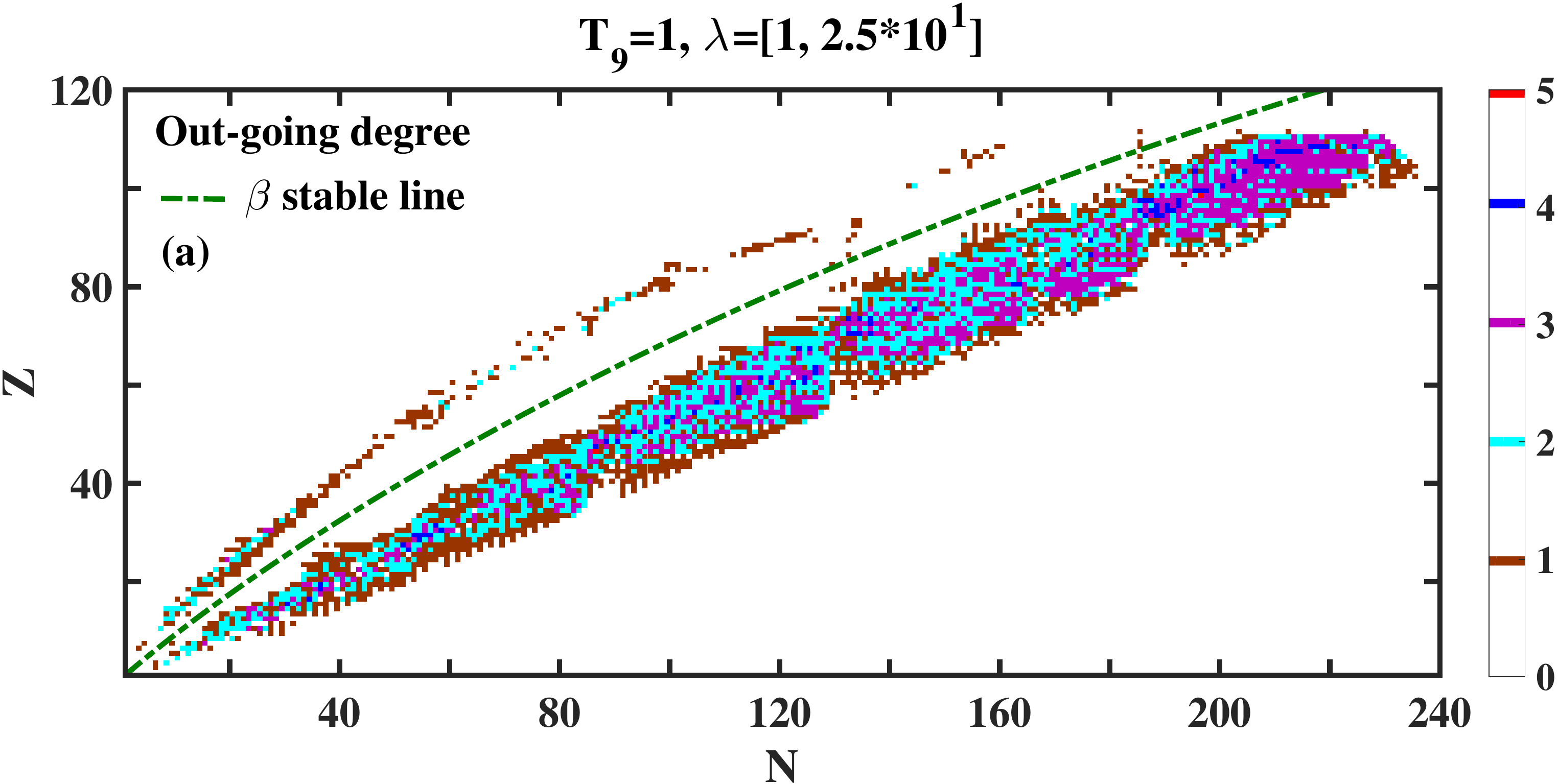}{}
   \includegraphics[width=8.0cm]{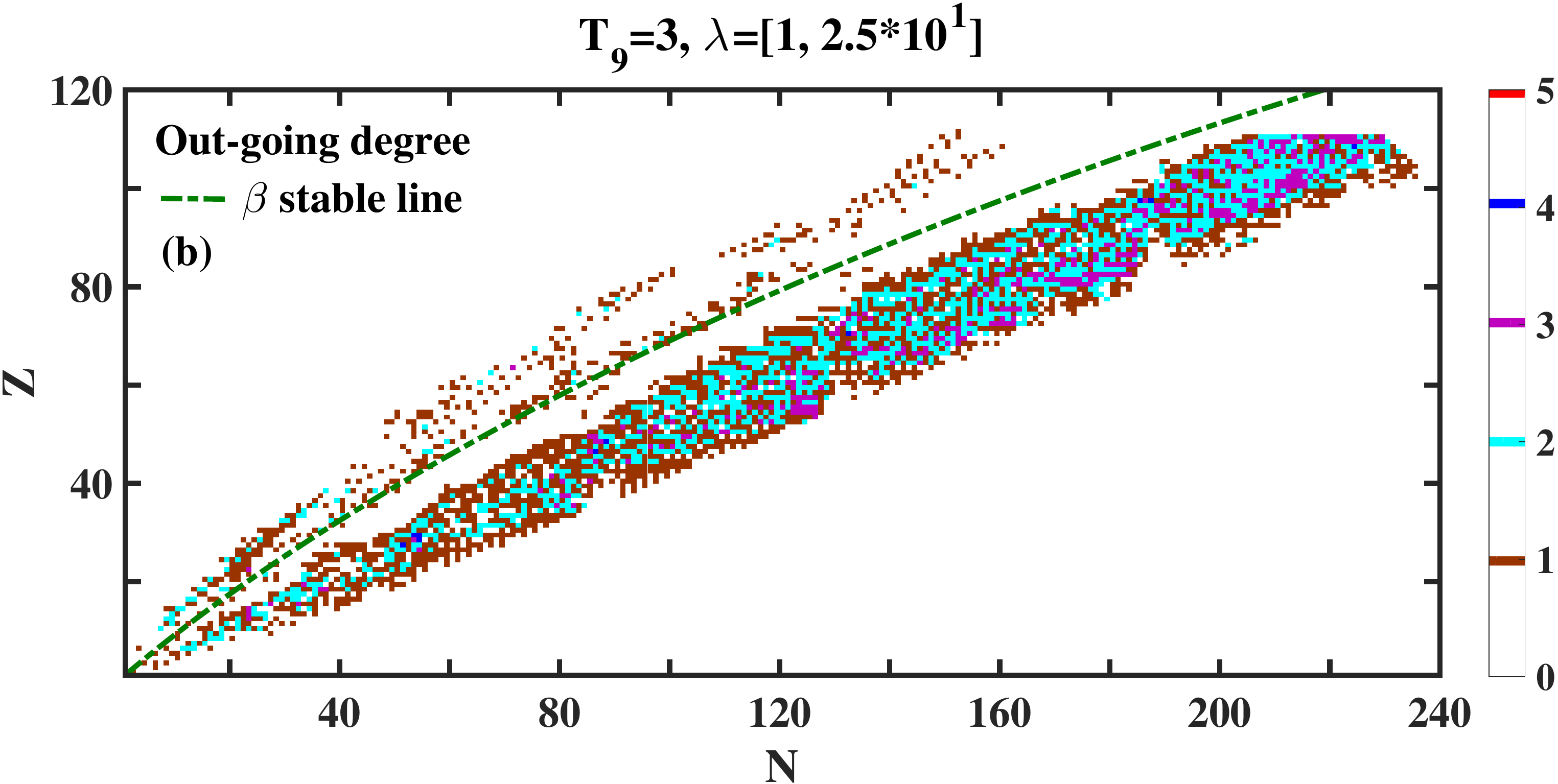}{}
   \includegraphics[width=9.2cm]{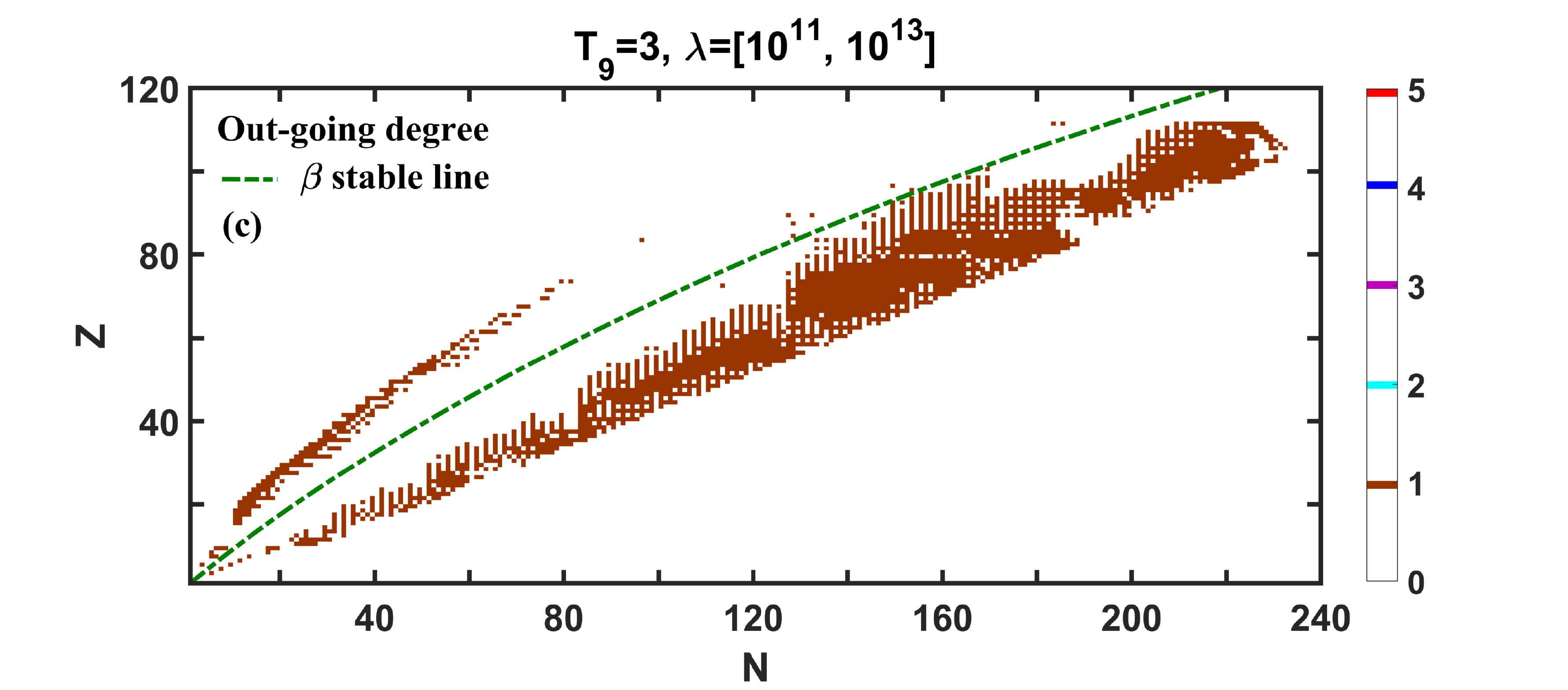}{} 
   \caption{ Out-going degree distributions of $R$-layer on a $N$-$Z$ plane in different windows of reaction rate ($\lambda$) and  temperature ($T_9$). 
   (a)  $\lambda = [10^{0} , 2.5\times10^{1}]$ and $T_{9}=1$; (b)  $\lambda = [10^{0} , 2.5\times10^{1}]$ and  $T_{9} = 3$;  (c): $\lambda = [10^{11} , 10^{13}]$ and $T_{9} = 3$.    The out-going degrees are calculated for each nucleus and the values are indicated by color and calibrated in the right border of the figure. $x$- and $y$-axis represent the number of neutrons ($N$) and  protons ($Z$)  of that nucleus. The $\beta$ stable line is also plotted in figure.}
\label{nuchart}
\end{figure}

It  shows that the central values of the peaks are $\lambda = 10^{0.6454}$ for $T_{9} = 1$, $\lambda = 10^{0.6714}$ and $10^{11.418}$ for $T_{9} = 3$, respectively. For the first peak at three  temperatures, the positions of peak vary not much, however, for the second peak which emerges at $T_{9} = 3$, the position of peak is much larger.
In the directed-weighted $R$-layer network, the direction is from reactant to product and the weight represents the capability of reactant to product. Therefore, the reaction-rate distribution  illustrates the attribute of the reactants, and the difference of  reaction rate  might express the different reactions of nuclei which can be indicated by the difference of the out-going degrees of nuclei in the network. There is a hypothesis that the reactions with $\lambda \leq 10^{-18}$ is unable to occur because these reactions will not change anything within  time scale of astrophysical process. In this work, we focus on the regions of $\lambda=[10^{0},2.5\times10^{1}]$ for $T_{9} = 1$,   $\lambda = [10^{0},2.5\times10^{1}]$ and  $\lambda = [10^{11},10^{13}]$ for $T_{9} = 3$, respectively, to study the nuclei's distribution with the number of nuclei  more than 80 percents of total mounts. Taking the advantage of the directed-weighted $R$-layer network, we compute the nuclei's out-going degree in the selected reaction rate's region, as  shown in the nuclei's chart. Figure~\ref{nuchart} demonstrates the out-going degree distributions of the nuclei in the regions of $\lambda = [10^{0}, 2.5\times10^{1}]$ for $T_{9} = 1$ as well as $\lambda = [10^{0},2.5\times10^{1}]$ and $\lambda = [10^{11} , 10^{13}]$ for $T_{9} = 3$. The nuclei in the first region of $\lambda = [10^{0}, 2.5\times10^{1}]$  display rich out-going degrees which have a maximum being equal to 5. However, all nuclei in the second region show the same out-going degrees being equal to 1. Such phenomenon reflects that the nuclei in the region of $\lambda = [10^{0},2.5\times10^{1}]$ at $T_{9} = 1$ and $T_{9} = 3$ can participate in multiple reactions, but the nuclei in the region of $\lambda = [10^{11},10^{13}]$ at $T_{9}=3$ just take part in one kind of reaction. The nuclei around the $\beta$ stable line emerge in the first peak region at $T_{9}$ = 3 but it does not appear at $T_{9}$=1. By comparing the nuclei in the regions of two peaks on the nuclear landscape at $T_{9}=3$, we find that there are differences in nuclei's distribution which implies that they may participate in different reactions. 

\begin{figure*}[!htbp]
    \centering
   \includegraphics[width=8.6cm]{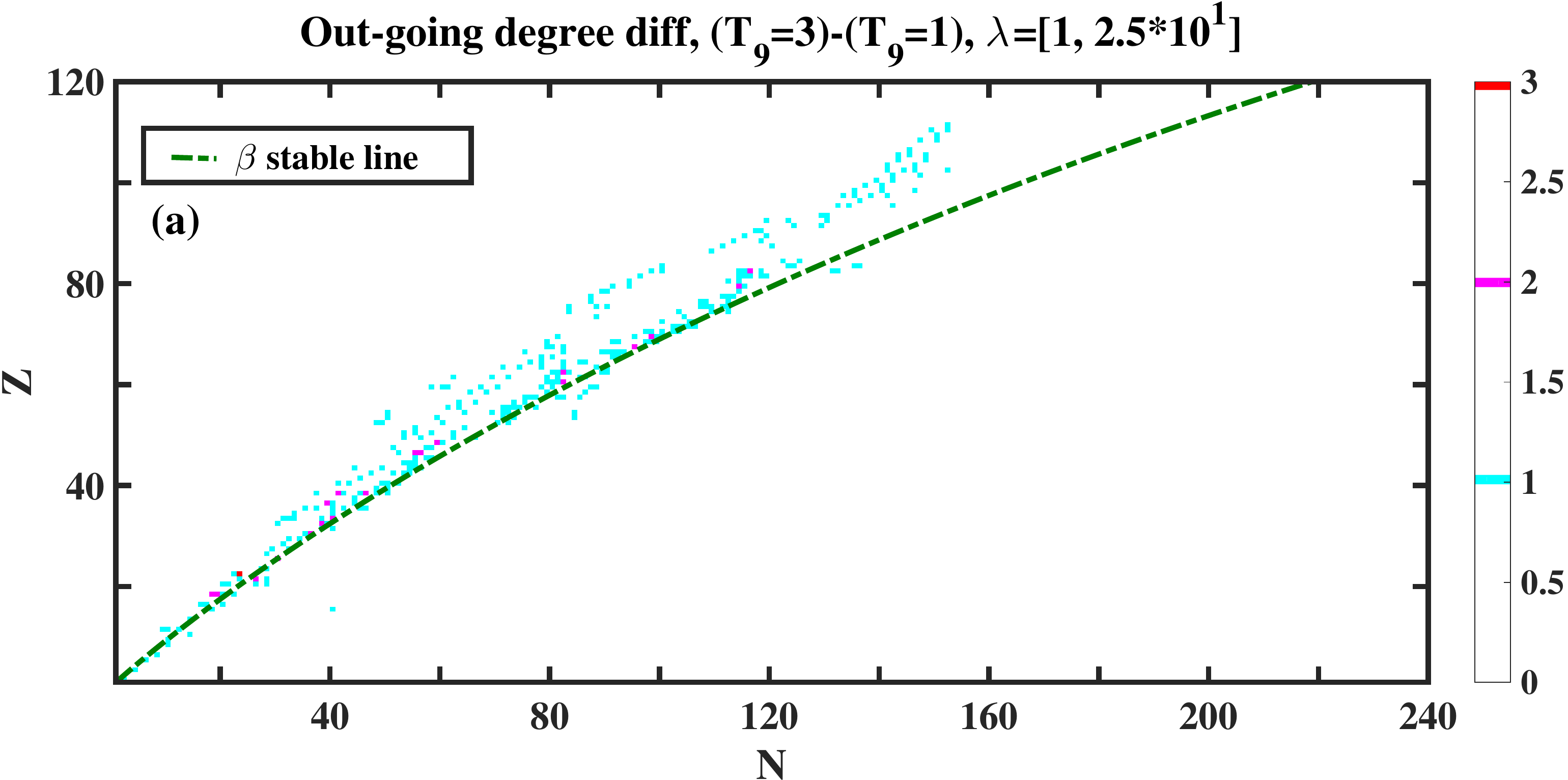}{}
   \includegraphics[width=8.6cm]{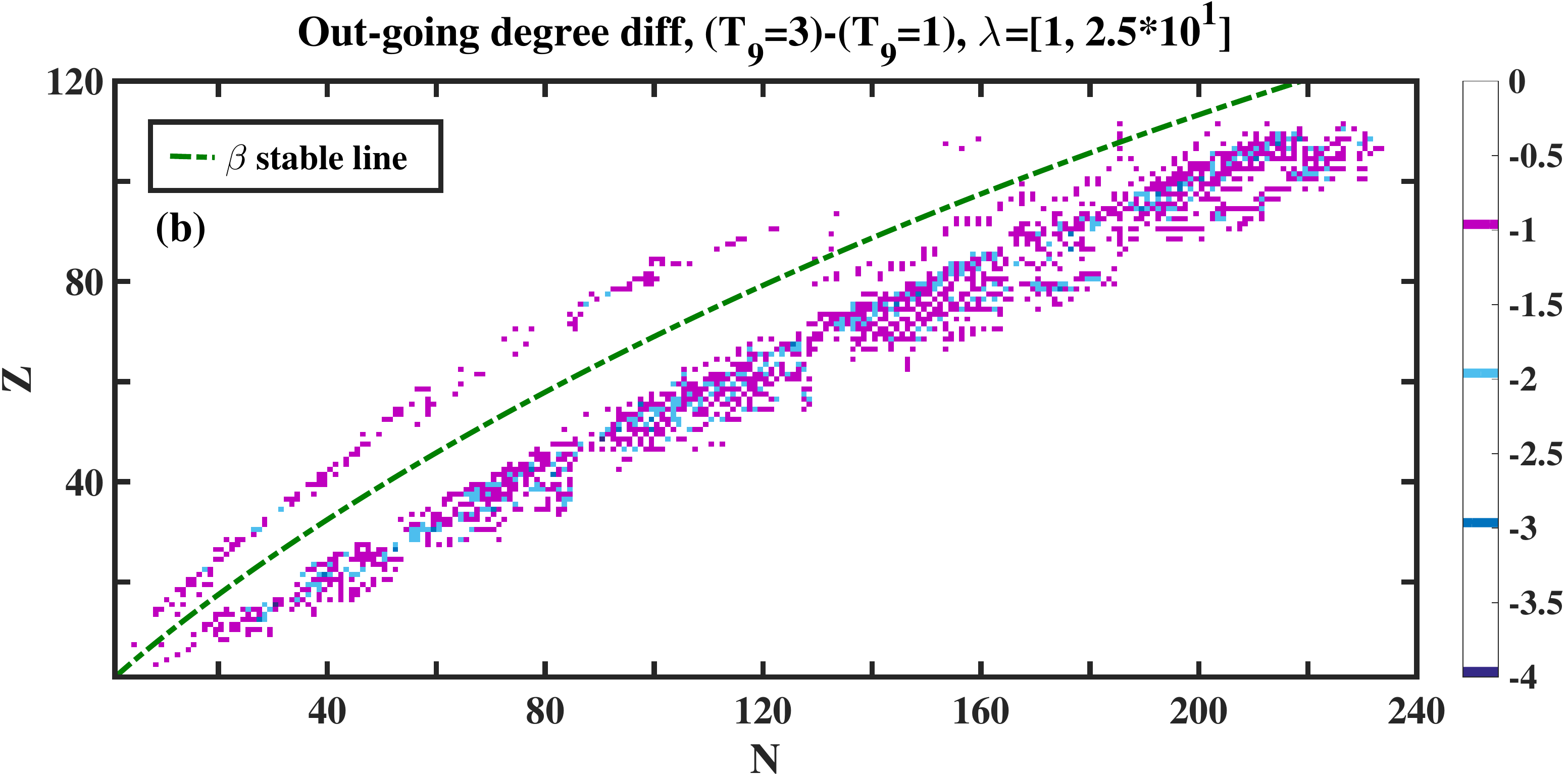}{} 
   \includegraphics[width=8.6cm]{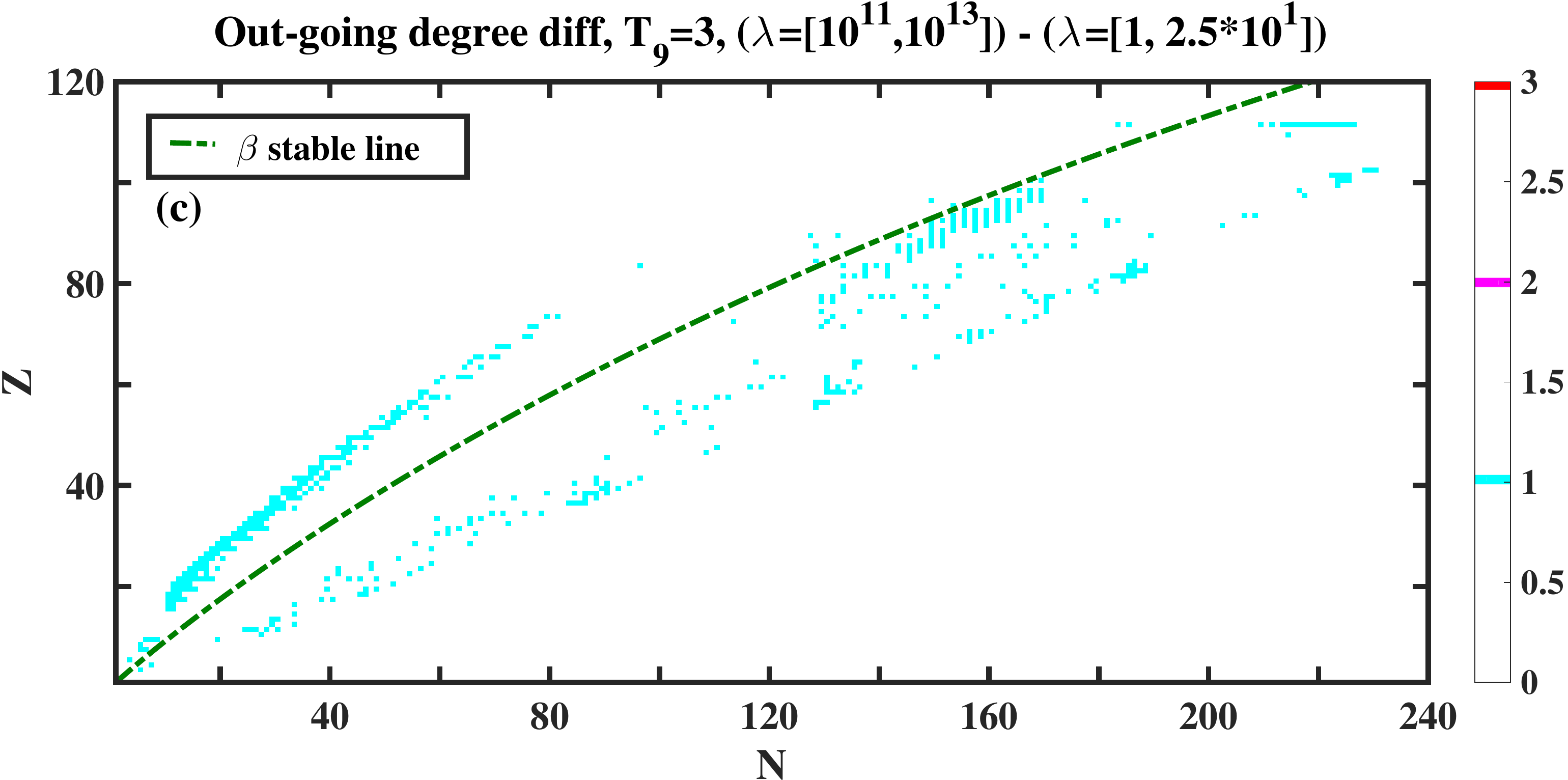}{}    
   \includegraphics[width=8.6cm]{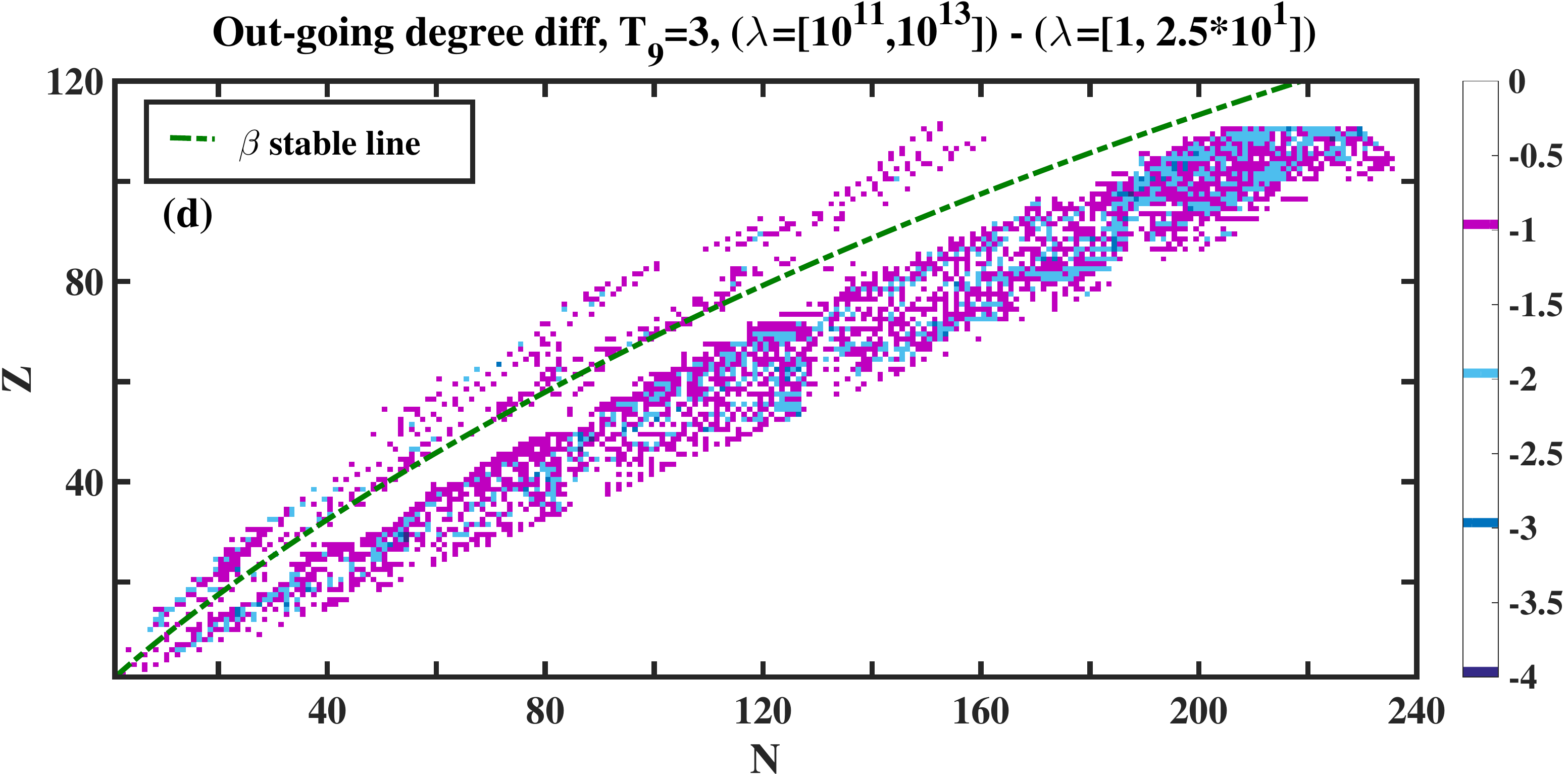}{}    
    \caption{Difference of out-going degrees between different regions of $\lambda$  and $T_9$. (a) positive degree difference between $T_{9} = 3$ and $T_{9} = 1$ in the region of  $\lambda = [10^{0}, 2.5\times10^{1}]$; (b) same as (a) but for the part of  negative degree difference; 
    (c) positive degree difference between  the regions of   $\lambda = [10^{11}, 10^{13}]$ and  $[10^{0}, 2.5\times10^{1}]$ at $T_{9} = 3$;
    (d) same as (c) but for the part of negative degree difference.}
    \label{Diff}
\end{figure*}

Furthermore, in order to demonstrate the influences of temperature and reaction rates, we perform the degree differences between Fig.~\ref{nuchart}(b) and (a), and plot them in upper panels of Fig.~\ref{Diff}.  Figure~\ref{Diff}(a)  shows the part of positive degree difference and (b)  shows the part of negative degree difference.   Figure~\ref{Diff}(a) indicates that the nuclei around the $\beta$ stable line emerge from $T_9$ = 1 to 
$T_9$ = 3  and stable nuclei more vulnerable to decay at high temperatures. Meanwhile, the elements in the neutron-rich region vanish and some proton-rich nuclei decline as shown in Fig.~\ref{Diff}(b).  In the same way, the degree differences between Fig.~\ref{nuchart}(c) and (b) are  divided into two panels:  Fig.~\ref{Diff}(c)  shows the part of positive degree difference and Fig.~\ref{Diff}(d)  shows the part of negative degree difference. From those two panels, small portions of proton-rich and neutron-rich nuclei are produced as plotted by all markers in Fig.~\ref{Diff}(c) at higher nuclear reaction rates, i.e $\lambda = [10^{11}, 10^{13}]$, but the nuclei around the $\beta$ stable line, and dominant neutron-rich nuclei as well as  some proton-rich nuclei disappear as plotted by all markers in Fig.~\ref{Diff}(d). These phenomena evidenced that neutron-rich and proton-rich nuclei can occur reactions with high reaction rates.

All reaction types in astrophysical investigations are $(n,\gamma), (n,p), (n,\alpha), (p,\gamma), (p,n), (p,\alpha), (\alpha,\gamma), (\alpha,n), (\alpha,p)$, 
$(\gamma,n), (\gamma,p), (\gamma,\alpha)$, $\beta^{+}$ and $\beta^{-}$. The main reaction types in the $R$-layer network, by using of the reactant-product method, include the reactions $(\gamma,n), (\gamma,p), (\gamma,\alpha)$, $\beta^{+}$ and $\beta^{-}$. Previously, we systematically introduced the nuclei's distribution on the nuclei chart which can react at different temperatures or reaction rates. In order to  explain  concretely the reaction type which nuclei are involved, we specify the types of reactions that each nucleus can participate in. Table I shows the detailed information of each reaction type in the $R$-layer, where $\delta A$ and $\delta Z$ are the difference in mass and proton numbers of the reactants and products, respectively.
\begin{table}[!htbp]
\label{types}
\caption{The classification of reaction types depending on the difference of mass number and proton number between reactant and production.}
\begin{tabular}{ccc}
\hline
\toprule
Reaction type& $\delta$A& $\delta$Z\\
\midrule
\hline
($\gamma$ , n)& 1& 0\\
($\gamma$ , p)& 1& 1\\
($\gamma$ , $\alpha$)& 4 & 2\\
$\beta^{+}$ & 0 & 1\\
$\beta^{-}$ & 0 & -1\\
\hline
\bottomrule
\end{tabular}
\end{table}

\begin{figure}[!htbp]
    \centering
  \includegraphics[width=8.6cm]{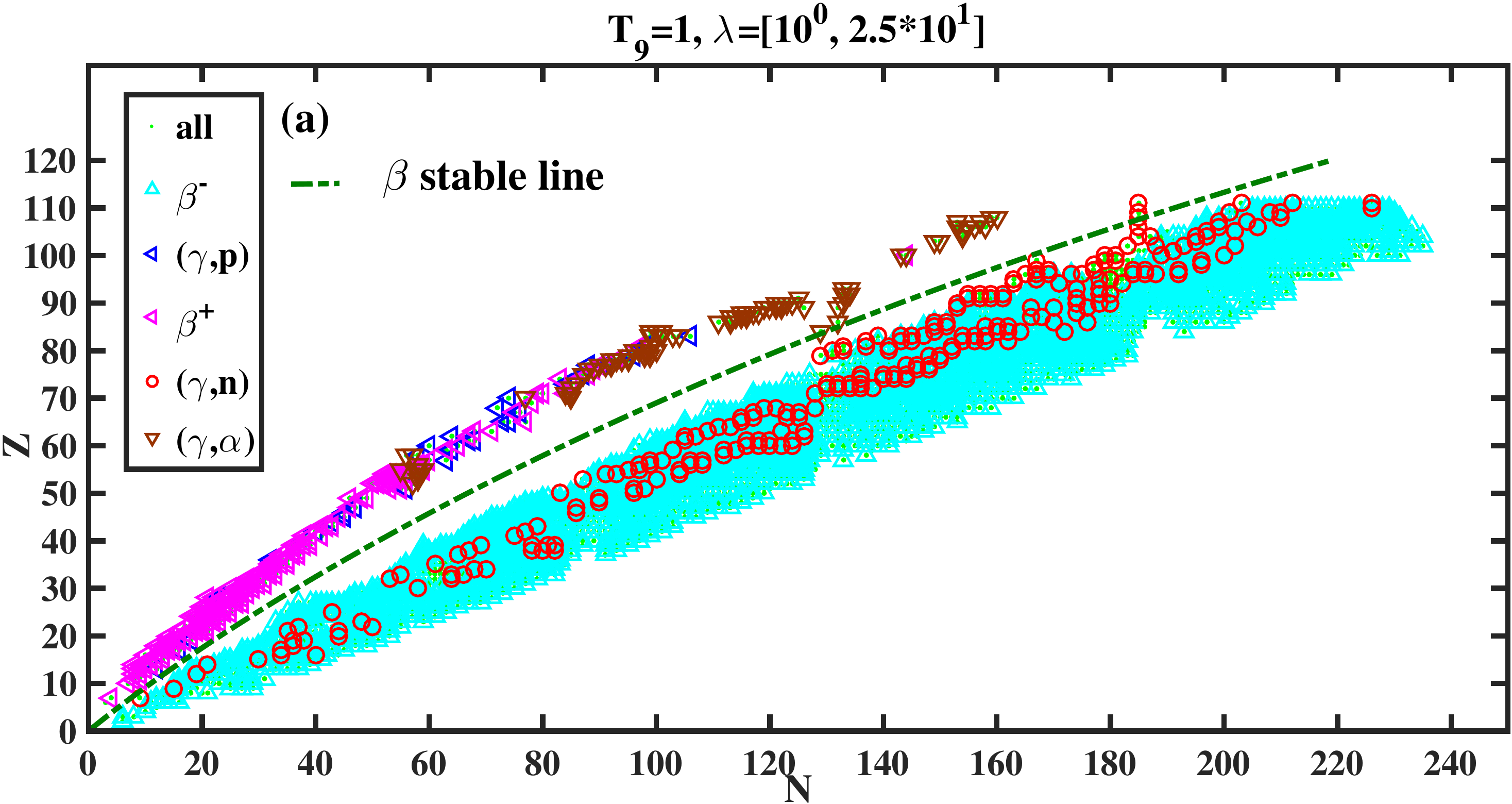}{}
   \includegraphics[width=8.6cm]{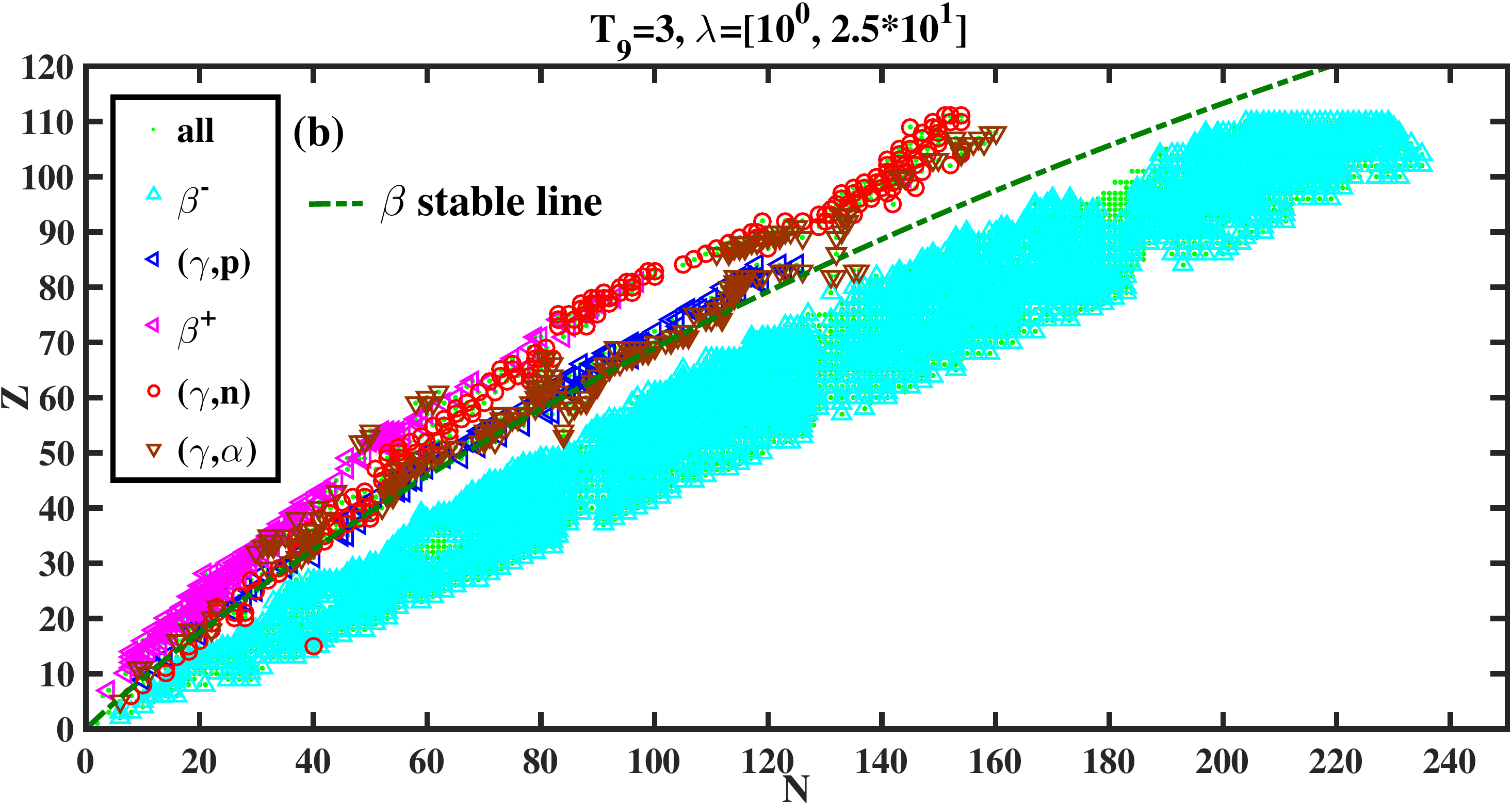}{} 
   \includegraphics[width=8.6cm]{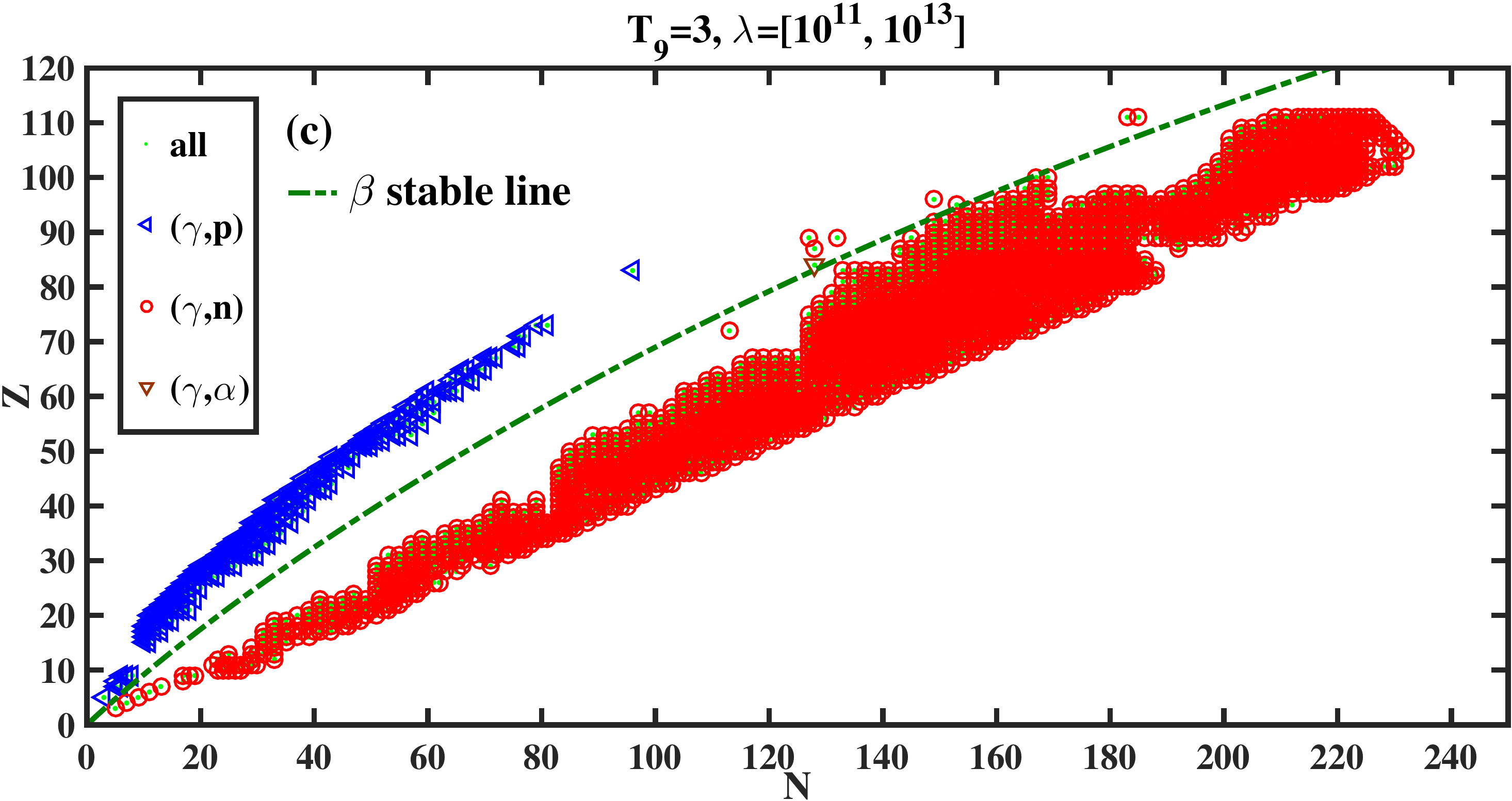}{}    
    \caption{ Nuclei's distributions of the $R$-layer on a $N$-$Z$ plane in different windows of  reaction rate ($\lambda$) and  temperature ($T_9$). 
        (a):  $\lambda = [10^{0}, 2.5\times10^{1}]$ and  $T_{9} = 1$. (b):  $\lambda = [10^{0}, 2.5\times10^{1}]$ and  $T_{9} = 3$, and (c):  $\lambda = [10^{11}, 10^{13}]$ and $T_{9} = 3$.
        The different reaction types are indicated by different markers with colors.  }
    \label{ReacNum}
\end{figure}

 According to the classical method of nuclear reaction types in Tab. I, we conduct a statistical analysis of the reactions which each nucleus can take part in at different temperatures. The nuclei's distribution of all reaction types in the nuclear chart is shown  in Fig.~\ref{ReacNum}. Different colors represent different nuclear reactions, and  more detailed information can be found in Fig.~\ref{ReacNum}. In  Fig.~\ref{ReacNum}(a) and (b),  most of the nuclei in the reaction rate region of $[10^{0},2.5\times10^{1}]$ can participate in  multiple reactions (i.e. some nuclei can participate in other reactions which not shown in Tab. I). For instance, these nuclei may decay multiple neutrons, protons or heliums. By comparing Fig.~\ref{ReacNum}(a) with (b),  we found that the accessorial nuclei mainly take part in  reactions of $(\gamma,p)$ and $(\gamma,\alpha)$, and these reactions are responsible for the emergence of the nuclei around the $\beta$ stable line at $T_9 =3$.  On the other hand, it  illustrates  that the nuclei which take part in $(\gamma,n)$ reaction at $T_{9}=1$ have a significantly difference  with respect to that at $T_{9} = 3.$
 
For the nuclei's distribution at the same temperature $T_{9} = 3$, the nuclei within two regions of nuclear reaction rates have   different ($N,Z$) distributions shown in Fig.~\ref{ReacNum}(b) and (c). The nuclei in neutron-rich region are more likely to have $(\gamma,n)$ reaction at higher temperature which can be attributed by the high neutron-proton ratio. Moreover, combining the nuclei's distributions of Fig.~\ref{ReacNum}(a) with those of Fig.~\ref{ReacNum}(b) and (c), we can see that the $(\gamma,n)$ reaction for proton-rich nuclei can occur at higher temperature the same as the neutron-rich nuclei. 
The nuclei, on the whole, in the first peak mainly involve $\beta$ decay, which are largely temperature independent in the JINA database.  And the nuclei in the second peak are due to  photodisintegration reactions, which become important at higher temperatures.
Specific numbers of the nuclei of each reaction type are shown in Tab. II, which  displays that the nuclei in the reaction rate region of $[10^{11},10^{13}]$ only take part in one reaction. According to the nuclei region and temperatures, the nuclear processes for neutron-rich and proton-rich nuclei are corresponding to r-process and rp-process. Both processes are the competition between $\beta$ decay and nucleon capture reactions.

\begin{table}[!htbp]
\label{alltypes}
\centering
\renewcommand\arraystretch{1}
\caption{The reaction types at $T_{9}=1$ and 3 with  different  $\lambda$ 
in the R-layer network.}
\begin{tabular}{lcccccc}
\hline
\toprule
Reaction & $T_{9} = 1$ & $T_{9} = 3$ & $T_{9} = 3$\\
 type & $\lambda = [10^{0}, 25]$ & $\lambda = [10^{0}, 25]$ & $\lambda = [10^{11}, 10^{13}]$\\
\midrule
\hline
($\gamma$, n)& 243& 144 & 2324\\
($\gamma$, p)& 35& 77 & 221\\
($\gamma$, $\alpha$)& 83& 162 & 1\\
$\beta^{+}$ & 162& 111 & \\
$\beta^{-}$ & 2487& 2232 & \\
all & 7331 & 5781 & 2546\\
\hline
\bottomrule
\end{tabular}
\end{table}

In summary, we constructed a new nuclear reaction-rate weighted multi-layer network based on the substrate-product method, which consists of all nuclei and reactions in the JINA REACLIB database. The nuclear reaction networks are set up with four layers of $N$-layer, $P$-layer, $H$-layer and $R$-layer, depending on the reactants of the neutron, proton, $^4$He or the reminder nuclei, respectively. Our special focus is on the $R$-layer reaction network since it has rich topological features. It is found that the weights of the nuclear reaction network display a heterogeneous distribution. Interestingly, with the increase of the $T_{9}$, the reaction rates of the $R$-layer network exhibit a transition from unimodal distribution  around $\lambda =   [10^{0}, 2.5\times10^{1}]$ to bimodal distribution with  one peak around $\lambda =   [10^{0}, 2.5\times10^{1}]$ and another around $\lambda = [10^{11}, 10^{13}]$. Based on the analysis of the nuclei within the bimodal peaks of nuclear reaction rates, we found that the nuclei within the first peak at $T_{9} = 1$ and $T_{9} = 3$ have a complicated out-going degree  in comparison with the nuclei in second peak at $T_{9}$ = 3, in which the out-going degree is equal to one. By classifying  all reactions in the $R$-layer at the same temperature $T_{9} = 3$, we found that the nuclei in the region of two nuclear reaction rates locate at very different $(N, Z)$ positions.
By comparing out-going degree distribution of nuclei at different temperatures of $T_{9} = 1$ and  3,  it is found that the  nuclei mainly take part in  reactions of $(\gamma,p)$ and $(\gamma,\alpha)$ at $T_{9} = 3$ which are responsible for the emergence of the nuclei around the $\beta$-stable line.  With the nuclear reaction rates approaching $\lambda = [10^{11}, 10^{13}]$, the nuclei around the $\beta$ stable line as well as the $\beta$ decays fade away, and 
the nuclei in neutron-rich region are mostly have only $(\gamma,n)$ reaction, which becomes the main reason of the instability of neutron-rich nuclei. 

We emphasize that the current work is not for making a specific diagnosis nucleus by nucleus, but for demonstrating a kind of big-data statistical  analysis of the nuclear chart.
In addition, caution should be taken because of the lack of some exotic nuclear reaction data and the extrapolation uncertainty of reaction rates in the current database-driven analysis. It is expected  that more and more unstable nuclear reaction and decay data will be accumulated with the application of radioactive ion beam accelerator, which will certainly further improve the network analysis.
Overall, the present work sheds light on that a novel way for 
the topological structure analysis of nuclear reaction network at  different stellar temperatures by a marriage of a known nuclear reaction-rate database to the knowledge of complex network science.  
 
{\it Acknowledgments}: This work is supported by the National Natural Science Foundation of China under Contracts Nos. 11890714, 11421505, 11875133 and 11075057, the National Key R$\&$D Program of China under Grant No. 2018YFB2101302, the Key Research Program of Frontier Sciences of the CAS under Grant No. QYZDJ-SSW-SLH002,   and the Strategic Priority Research Program of the CAS under Grant No XDB34030200.

\end{CJK*}
\end{document}